\documentclass{JAC2003}

\usepackage{graphicx}
\usepackage{booktabs}
\usepackage{multicol}
\usepackage{amsmath}
\usepackage{url}

\begin{document}
\title{\uppercase{Models of the magnetic field enhancement at pits}}

\author{Takayuki Kubo\thanks{kubotaka@post.kek.jp}, \\
KEK, High Energy Accelerator Research Organization, 1-1 Oho, Tsukuba, Ibaraki 305-0801 Japan}

\maketitle

\begin{abstract}
Models of the magnetic field enhancement at pits are discussed. 
In order to build a model of pit, parameters that characterize a geometry of edges of pit, 
such as a curvature radius and a slope angle of edge, should be included, 
because the magnetic field is enhanced at an edge of a pit. 
A shape of the bottom of the pit is not important, because the magnetic field attenuates at the bottom. 
The simplest model of the pit is given by the two-dimensional pit with a triangular section. 
The well-like pit, which is known to many researchers, is a special case of this model. 
In this paper, idea and methods to analytically evaluate the magnetic field enhancement factor of such models are mentioned in detail. 
The well-like pit is considered as an instructive exercise, where the famous results by Shemelin and Padamsee are reproduced analytically. 
The triangular-pit model, which is practically important for studies of the quench of SRF cavity, are discussed in detail. 
Comparisons between the prediction of the triangular-pit model and the vertical test results are also shown. 
\end{abstract}

\section{Introduction}

The magnetic field locally enhanced at a pit on the surface of the superconducting radio-frequency (SRF) cavity 
is thought to be a cause of quenches.  
The enhanced magnetic field at a pit is written as  
\begin{eqnarray}
H({\bf r})=\beta({\bf r}) H_0 \hspace{0.8cm} (0\le \beta({\bf r}) \le \beta^*) \,, 
\end{eqnarray}
where $H_0$ is a surface magnetic field of the cavity, 
$\beta({\bf r})$ is a magnetic field enhancement (MFE) factor at a position ${\bf r}$,  
and $\beta^*$ is the maximum value of $\beta({\bf r})$ along the pit.   
To reveal the relation between the MFE factor and the geometry of pit is the challenge that should be solved in order for a better understanding of the quench at pits.

In order to build a model of pit, parameters that characterize a geometry of edges of pit, 
such as a curvature radius and a slope angle of edge, should be included, 
because the magnetic field is enhanced at an edge of a pit. 
A shape of the bottom of the pit is not important, because the magnetic field attenuates at the bottom. 
The simplest model of the pit is given by the two-dimensional pit with a triangular section~\cite{kubo}. 
The well-like pit, which is known to many researchers, is a special case of this model.

In following sections, idea and methods to analytically evaluate the magnetic field enhancement factor of such models are mentioned in detail. 
First a neccesary mathematical-technique, the method of conformal mapping, is reviewed.  
Then the well-like pit model is considered as an instructive exercise, 
where we see the magnetic field is enhanced at an edge of a pit, and confirm the magnetic field attenuates at the bottom. 
The functional form of the MFE factor of the well-like pit derived by Shemelin and Padamsee~\cite{shemelin} are also reproduced.  
Then the triangular pit model, which is practically important for studies of the quench of SRF cavity, is discussed. 
Comparisons between a prediction of the model and vertical tests results are also shown.

\section{Method of conformal mapping} 

\begin{figure}[tb]
   \begin{center}
   \includegraphics[width=0.8\linewidth]{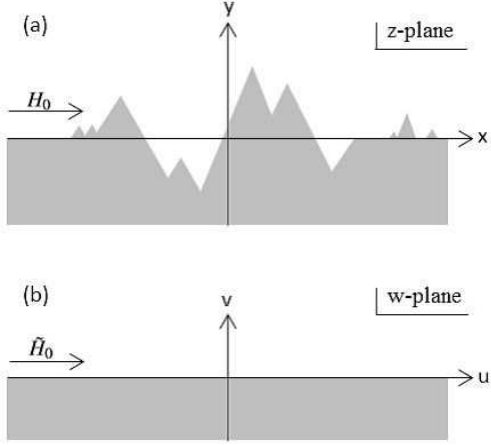}
   \end{center}\vspace{-0.6cm}
   \caption{
Two-dimensional domains on complex planes.   
Gray regions represent a superconductor in the Meissner state. 
   }\label{figure1}
\end{figure}

Let us consider a two-dimensional domain shown in Fig~\ref{figure1}(a). 
In order to derive the magnetic field distribution, 
the Maxwell equations of a two-dimensional magnetostatics should be solved. 
This problem is equivalent to that of finding an appropriate holomorphic function called the complex potential on a complex plane. 
The magnetic field is given by the derivative of the complex potential $\Phi(z)$, 
\begin{eqnarray}
H_x(x,y) - i H_y(x,y)  = -\frac{d\Phi(z)}{dz}  \,,    \label{eq:HxHy1}
\end{eqnarray}
where $z=x+iy$. 
The method of conformal mapping provides a powerful tool to compute the right hand side of Eq.~(\ref{eq:HxHy1}).

Suppose a domain on the $z$-plane shown in Fig~\ref{figure1}(a) is mapped into 
a simple domain on the $w$-plane shown in Fig~\ref{figure1}(b) by a Schwarz-Christoffel transformation, 
\begin{eqnarray}
z= F(w) = K_1 \int_0^w \!\! f(w')  dw' + K_2 \, ,\label{eq:SC}  
\end{eqnarray}
where $K_1$ and $K_2$ are constants, and $f(w)$ is a holomorphic function. 
Then orthogonal sets of field lines on one plane are transformed into those on another plane, and  
the complex potential on the $z$-plane, $\Phi(z)$, is related to that on the $w$-plane, $\tilde{\Phi}(w)$, as
\begin{eqnarray}
\Phi(z)=\Phi(F(w))=\tilde{\Phi}(w)  \,.   
\end{eqnarray}
Thus Eq.~(\ref{eq:HxHy1}) becomes
\begin{eqnarray}
H_x - i H_y = -\frac{d\tilde{\Phi}(w)}{dw}\frac{dw}{dz} = -\frac{d\tilde{\Phi}/dw}{dz/dw} = -\frac{\tilde{\Phi}'(w)}{K_1 f(w)} \,.   \label{eq:HxHy2}
\end{eqnarray}
Since the complex potential $\tilde{\Phi}(w)$, 
which is a holomorphic function on the $w$-plane and is related to the magnetic field on the $w$-plane as $H_u -iH_v=\tilde{\Phi}'(w)$, 
can be written as $\tilde{\Phi}(w)= - \tilde{H}_0 w = -K_1 H_0 w$,   
Eq.~(\ref{eq:HxHy2}) is reduced to 
\begin{eqnarray}
H_x - i H_y  =  \frac{H_0}{f(w)}  \,.   \label{eq:HxHy3}
\end{eqnarray}
The MFE factor, which is defined by the ratio of the magnetic field strength at a position to that far from the pit, 
is given by
\begin{eqnarray}
\beta = \frac{\sqrt{H_x^2 +H_y^2}}{H_0} = \frac{1}{|f(w)|} \,. \label{eq:beta}
\end{eqnarray}
This is the general formula for the MFE factor when the map is given by a Schwarz-Christoffel transformation. 

\section{Well-like pit} 

In this section we discuss a two-dimensional well. 
Although this model is not practically useful, it provides an instructive exercise. 
First we consider a well with sharp edges as an effective model of the well with round edges. 
Then the MFE factor of the well with round edges are estimated in a naive way. 

\subsection{Well-like pit with sharp edges}

\begin{figure}[t]
   \begin{center}
   \includegraphics[width=0.8\linewidth]{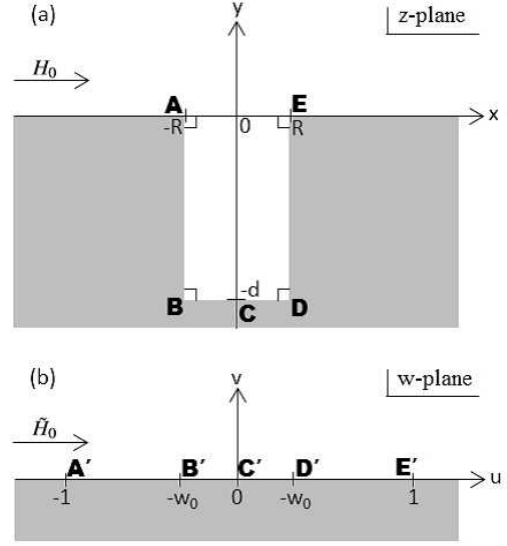}
   \end{center}\vspace{-0.6cm}
   \caption{
A two-dimensional well with sharp edges on (a) the $z$-plane and (b) the $w$-plane. 
Gray regions represent to a superconductor in the Meissner state, 
$R$ is half a width and $d\,(\gg R)$ is a depth of the well. 
   }\label{figure2}
\end{figure}

Let us consider a two-dimensional well with sharp edges shown in Fig.~\ref{figure2}. 
Note that $d\gg R$ is assumed for simplicity. 
The map is given by Eq.~(\ref{eq:SC}), where 
\begin{eqnarray}
f(w) &=& \Bigl( \frac{w^2-1}{w^2-w_0^2} \Bigr)^{\frac{1}{2}}  \,. \label{eq:f_well} 
\end{eqnarray}
The constants $K_1$, $K_2$ and $w_0$ are determined by conditions that C', D' and E' on the $w$-plane are mapped into C, D and E on the $z$-plane: 
\begin{eqnarray}
K_1 \simeq \frac{2R}{\pi} \,, \hspace{0.5cm} 
K_2 = -id \,, \hspace{0.5cm}
w_0 \simeq  e^{-\frac{\pi}{2}\frac{d}{R}}  \,.
\end{eqnarray}
Substituting Eq.~(\ref{eq:f_well}) into Eq.~(\ref{eq:beta}), we find
\begin{eqnarray}
\beta_{\rm well} 
=\frac{1}{|f(w)|}= \Bigl| \frac{w^2-w_0^2}{w^2-1} \Bigr|^{\frac{1}{2}} \, . \label{eq:beta_well_1}
\end{eqnarray}
Thus the MFE factor at A, B, C, D or E is given by 
\begin{eqnarray}\label{eq:betaABCDE}
  \begin{cases}
    \beta_{\rm well}({\rm A}) = \beta_{\rm well}({\rm E}) = \infty       & (w=\pm 1)   \,, \\
    \beta_{\rm well}({\rm B}) = \beta_{\rm well}({\rm D}) = 0            & (w=\pm w_0) \,, \\
    \beta_{\rm well}({\rm C}) = w_0 \simeq e^{-\frac{\pi}{2}\frac{d}{R}} & (w=0) \,.
  \end{cases}
\end{eqnarray}
The MFE factor diverges at the edges of the pit (A and E), 
vanishes at the concave corners (B and D), 
and exponetially attenuates at the bottom (C). 

In order to evaluate the MFE factor at an arbitrary $z$, however, $f(w)$ in Eq.~(\ref{eq:beta_well_1}) should be expressed in terms of $z$. 
This task is approximately achieved by focusing on a small region around a corner. 
Let us consider the vicinity of the corner E, which correspond to the vicinity of E' on the $w$-plane, 
which can be written as $w=1+\epsilon$ with $\epsilon \ll 1$. 
Then the function $f(w)$ is approximated as $f(w)\simeq (2\epsilon)^{\frac{1}{2}}$ at $w\sim 1$, 
and thus Eq.~(\ref{eq:SC}) is computed as $z=F(w)=R + (K_1/3)f^3(w)$, or
\begin{eqnarray}
f(w) = \biggl( \frac{3\pi}{2}\frac{z-R}{R} \biggr)^\frac{1}{3} \, . \label{eq:fz}
\end{eqnarray}
Substituting Eq.~(\ref{eq:fz}) into Eq.~(\ref{eq:beta_well_1}), we find 
\begin{eqnarray}
\beta_{\rm well} (r) = \biggl( \frac{2}{3\pi} \frac{R}{r} \biggr)^\frac{1}{3} \, , \label{eq:beta_well_2}
\end{eqnarray}
where $r=|z-R|$ is a distance from the corner E. 
The same formula can be applied to the MFE factor around the corner A by replacing $r=|z-R|$ with $r=|z+R|$.  
Note that Eq.~(\ref{eq:beta_well_2}) reproduces $\beta_{\rm well}(A)=\beta_{\rm well}(E)=\infty$ ($r\to0$).

\subsection{Well-like pit with round edges}

%
\begin{figure}[t]
   \begin{center}
   \includegraphics[width=0.8\linewidth]{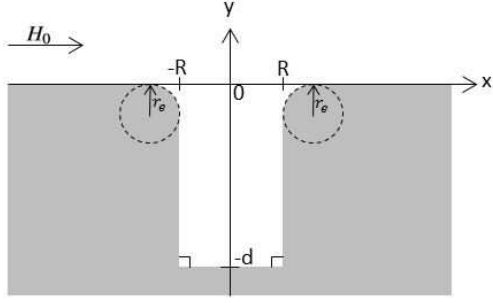}
   \end{center}\vspace{-0.6cm}
   \caption{
A two-dimensional well with round edges. 
A gray region represents a superconductor in the Meissner state,  
$R$ is half a width, $d\,(\gg R)$ is a depth of the well and $r_e$ is a round-edge radius. 
   }\label{figure3}
\end{figure}

Using Eq.~(\ref{eq:beta_well_2}), it is possible to estimate the maximum MFE factor for the well-like pit with round edges shown in Fig.~\ref{figure3}. 
As shown in the figure, at the domain $r=|z\pm R|\ge r_e$, the well-like pit with round edges has the same geometry with that with sharp edges. 
Thus the well-like pit with sharp edges can be regarded as an effective model of that with round edges, where a cutoff scale is given by $r_{\Lambda} = r_e$. 
Then we can naively estimate, 
\begin{eqnarray}
\beta^*_{\rm well} 
\sim \beta_{\rm well}(r_\Lambda) 
= P_{\rm well}^{\rm (naive)} \biggl( \frac{R}{r_e} \biggr)^\frac{1}{3} \, ,  \label{eq:beta_well_3}
\end{eqnarray}
where the coefficient is given by $P_{\rm well}^{\rm (naive)} = (2/3\pi)^\frac{1}{3}$. 
The functional form given by Eq.~(\ref{eq:beta_well_3}) corresponds to the result obtained by Shemelin and Padamsee~\cite{shemelin}. 
Note that the coefficient $P_{\rm well}^{\rm (naive)}$ usually yields a smaller value than exact value. 
A method to improve the estimate is found in Ref~\cite{kubo}.

\section{Triangular pit} 

As mentioned in the introduction, 
the simplest model of the pit is given by a two-dimensional pit with a triangular section. 
In a following, the similar method as the last section is applied to the triangular-pit model. 

\subsection{Pit with sharp edges} 

First we consider a pit with sharp edges shown in Fig.~\ref{figure4}. 
The MFE factor around the convex corners A or C can be computed in much the same way as above: 
\begin{eqnarray}
\beta_{\rm tri}(r) 
= \Bigl[  \frac{\sqrt{\pi} }{2\alpha(\alpha +1) \cos{\pi\alpha} \, \Gamma(\alpha)\Gamma(\frac{1}{2}-\alpha)} \frac{R}{r} \Bigr]^{\frac{\alpha}{1+\alpha}} \,. 
\label{eq:beta_tri_1}
\end{eqnarray}
where $r$ is a distance from the corner A or C.

\subsection{Pit with round edges} 
\begin{figure}[tb]
   \begin{center}
   \includegraphics[width=0.8\linewidth]{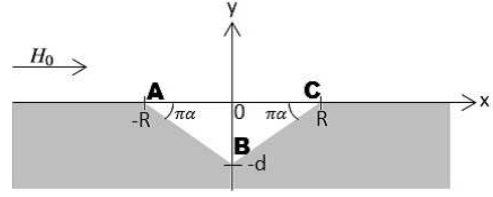}
   \end{center}\vspace{-0.6cm}
   \caption{
A two-dimensional triangular pit with sharp edges.  
A gray region corresponds to a superconductor in the Meissner state,  
$\pi \alpha\, (0< \alpha <1/2)$ is a slope angle and $R$ is half a width of the open mouth. 
Note that a relation between a depth, $d$, and half a width of the open mouth, $R$, is given by $d=R\tan\pi\alpha$. 
   }\label{figure4}
\end{figure}

\begin{figure}[tb]
   \begin{center}
   \includegraphics[width=0.8\linewidth]{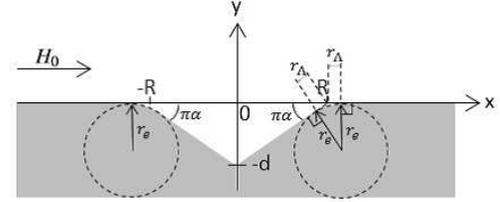}
   \end{center}\vspace{-0.6cm}
   \caption{
A two-dimensional triangular pit with round edges.  
A gray region corresponds to a superconductor in the Meissner state,  
$\pi \alpha\, (0< \alpha <1/2)$ is a slope angle, $R$ is half a width of the open mouth, 
and $r_e$ is a radius of the round edge. 
   }\label{figure5}
\end{figure}

The triangular pit with sharp edges shown in Fig.~\ref{figure4} can be regarded as an effective model of that with round edges shown in Fig.~\ref{figure5}, 
where a cutoff scale is given by $r_{\Lambda} = r_e (1-\cos\pi\alpha)/\sin\pi\alpha$.  
By substituting $r=r_{\Lambda}$ into Eq.~(\ref{eq:beta_tri_1}), 
we can estimate the maximum MFE factor of the model with round edges as 
\begin{eqnarray}
\beta^*_{\rm tri} 
\sim \beta_{\rm tri}(r_\Lambda) 
= P_{\rm tri}^{\rm (naive)}(\alpha) \biggl( \frac{R}{r_e} \biggr)^{\frac{\alpha}{1+\alpha}}  \, ,  \label{eq:beta_tri_2}
\end{eqnarray}
where the coefficient is given by
\begin{eqnarray}
\!\!\!\!P_{\rm tri}^{\rm (naive)}(\alpha) = 
\biggl[  
\frac{\sqrt{\pi} \tan\pi\alpha }
     {2\alpha(\alpha +1)(1\!-\!\cos\pi\alpha) \Gamma(\alpha)\Gamma(\frac{1}{2}\!-\!\alpha)} 
\biggr]^{\frac{\alpha}{1+\alpha}} \!\!\!\!\!.
\end{eqnarray}
Noted that $P_{\rm tri}^{\rm (naive)}$ is reduced to $P_{\rm tri}^{\rm (naive)} \to (2/3\pi)^\frac{1}{3} = P_{\rm well}^{\rm (naive)}$ at the limit $\alpha \to 1/2$, 
where the Euler's reflection formula was used. 
The coefficient $P_{\rm tri}^{\rm (naive)}$ yields a smaller value than exact value, in common with that of the well-like pit. 


\subsection{Improved formula} 

\begin{figure}[tb]
   \begin{center}
   \includegraphics[width=1\linewidth]{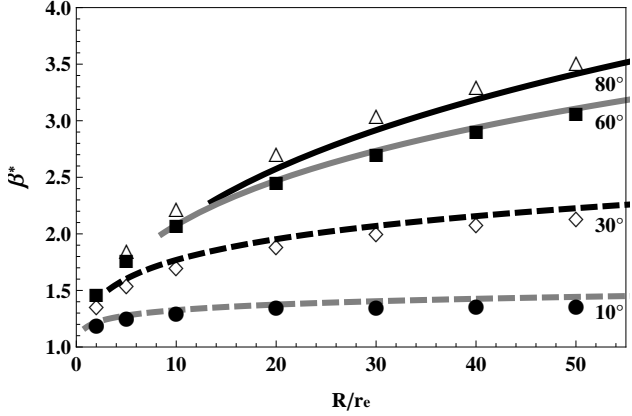}
   \end{center}\vspace{-0.6cm}
   \caption{
Agreements between the analytical formula given by Eq.~(\ref{eq:betaI}) with $n=6$ and numerical calculations by POISSON/SUPERFISH. 
The four curves show Eq.~(\ref{eq:betaI}) as functions of $R/r_e$ in applicable ranges~\cite{kubo}. 
The black line, gray line, black dashed line and gray dashed line correspond to 
slope angles $80^{\circ}$, $60^{\circ}$, $30^{\circ}$, and $10^{\circ}$ ($\alpha = 4/9$, $1/3$, $1/6$, and $1/18$), respectively. 
The four types of symbols show the maximum MFE factors for given pits calculated by POISSON/SUPERFISH, 
where the triangles, filled squares, rhombi, and filled circles correspond to 
slope angles $80^{\circ}$, $60^{\circ}$, $30^{\circ}$, and $10^{\circ}$, respectively. 
   }\label{figure6}
\end{figure}

The improved version of the formula can be obtained by extrapolating Eq.~(\ref{eq:beta_tri_1}) to $r\le r_{\Lambda}$. 
According to the complete results and discussions in Ref.~\cite{kubo}, $\beta_{\rm tri}^*$ is given by 
\begin{eqnarray}
\beta_{\rm tri}^* \,\, 
= P(\alpha) \, \biggl( \frac{R}{r_e} \biggr)^\frac{\alpha}{1+\alpha}
\,, \label{eq:betaI}
\end{eqnarray}
with
\begin{eqnarray} \label{eq:p}
\!\!\!\!\!\!\!\!\!P(\alpha) 
&=&\frac{1}{(n-1)!}\Bigl( 1+\frac{\gamma}{n}\Bigr) (2+\gamma) \cdots (n-1+\gamma) \nonumber\\  
&\times &
\biggl[  
\frac{\sqrt{\pi} \tan\pi\alpha }
     {2\alpha(\alpha +1)(1\!-\!\cos\pi\alpha) \Gamma(\alpha)\Gamma(\frac{1}{2}\!-\!\alpha)} 
\biggr]^\gamma \,, 
\end{eqnarray}
where $\gamma \equiv \alpha/(1+\alpha)$. 
The optimum degree of polynomial of extrapolation should be determined by experiments or numerical experiments.  
Fig.~\ref{figure6} shows agreements between $\beta_{\rm tri}^*$ with $n=6$ and numerical calculations by POISSON/SUPERFISH. 
The four curves correspond to Eq.~(\ref{eq:betaI}), and the four types of symbols correspond to numerical calculations by POISSON/SUPERFISH. 

Now we have the formula that yield the MFE factor of the pit with a triangular section. 
The next task is to confirm whether the triangular-pit model is a valid model or not. 

\section{Discussions} 

\subsection{Predictions of the model} 

Assuming that vortices start to penetrate into the defect-less superconductor at $B_v \,(\simeq 200\,{\rm mT})$~\footnote{Solving the Ginzburg-Landau (GL) equation, the superheating field  can be computed, and is given by $B_{\rm sh} \simeq 1.2 B_c$ at $T \simeq T_c$. In some literature, this GL result is applied to low temperature $T\ll T_c$ and the superheating field is given by $B_{\rm sh} \simeq 1.2 B_c = 240\,{\rm mT}$, but in this temperature region the GL theory becomes invalid. To evaluate $B_{\rm sh}$ at $T \ll T_c$, Eilenberger equations should be solved~\cite{gurevich}. In this paper, an empirical limit $\simeq 200\,{\rm mT}$ is adopted.} 
the achievable surface magnetic field without vortex dissipations under an existence of a pit is given by
\begin{eqnarray}
B_{\rm peak}^{\rm (pen)} =\frac{B_v}{\beta^*}= \frac{B_v}{P(\alpha)} \biggl( \frac{r_e}{R} \biggr)^{\frac{\alpha}{1+\alpha}} \,.  \label{eq:Hmax}
\end{eqnarray}
This equation is further reduced to the formula that describes the accelerating field at which vortices start to penetrate:  
\begin{eqnarray}
E_{\rm acc}^{\rm (pen)} = g^{-1} B_{\rm peak}^{\rm (pen)} = \frac{g^{-1} B_v}{P(\alpha)} \biggl( \frac{r_e}{R} \biggr)^{\frac{\alpha}{1+\alpha}} \,.  \label{eq:Epen}
\end{eqnarray}
where $g$ is a ratio of the peak magnetic field to the accelerating field for a given cavity-shape. 
A quench field $E_{\rm acc}^{\rm (quench)}$ is expected to be above the vortex penetration field, $E_{\rm acc}^{\rm (pen)}$, 
where normal conducting areas are formed due to heats of vortex dissipations, and a thermal-magnetic breakdown occurs.

\subsection{Comparisons with vertical test results}

\begin{figure}[t]
   \begin{center}
   \includegraphics[width=0.9\linewidth]{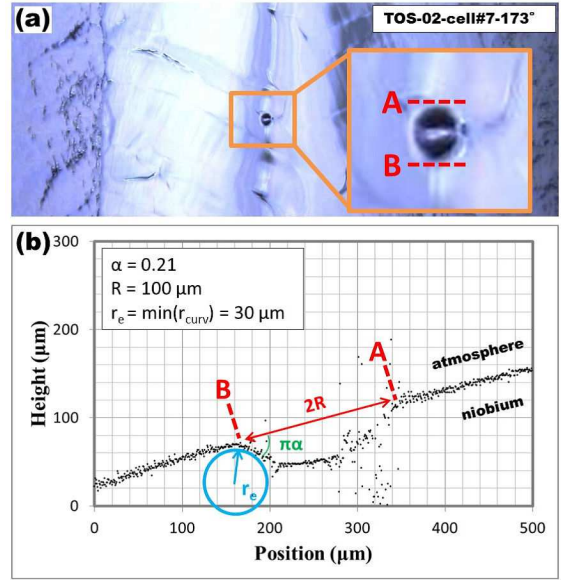}
   \end{center}\vspace{-0.6cm}
   \caption{
   A pit found at the surface of TOS-02 cavity. 
   (a) An optical image of the pit~\cite{yamamoto1}. 
   (b) Results of profilometry by a laser microscope. 
   }\label{figure7}
\end{figure}

\begin{figure}[t]
   \begin{center}
   \includegraphics[width=0.9\linewidth]{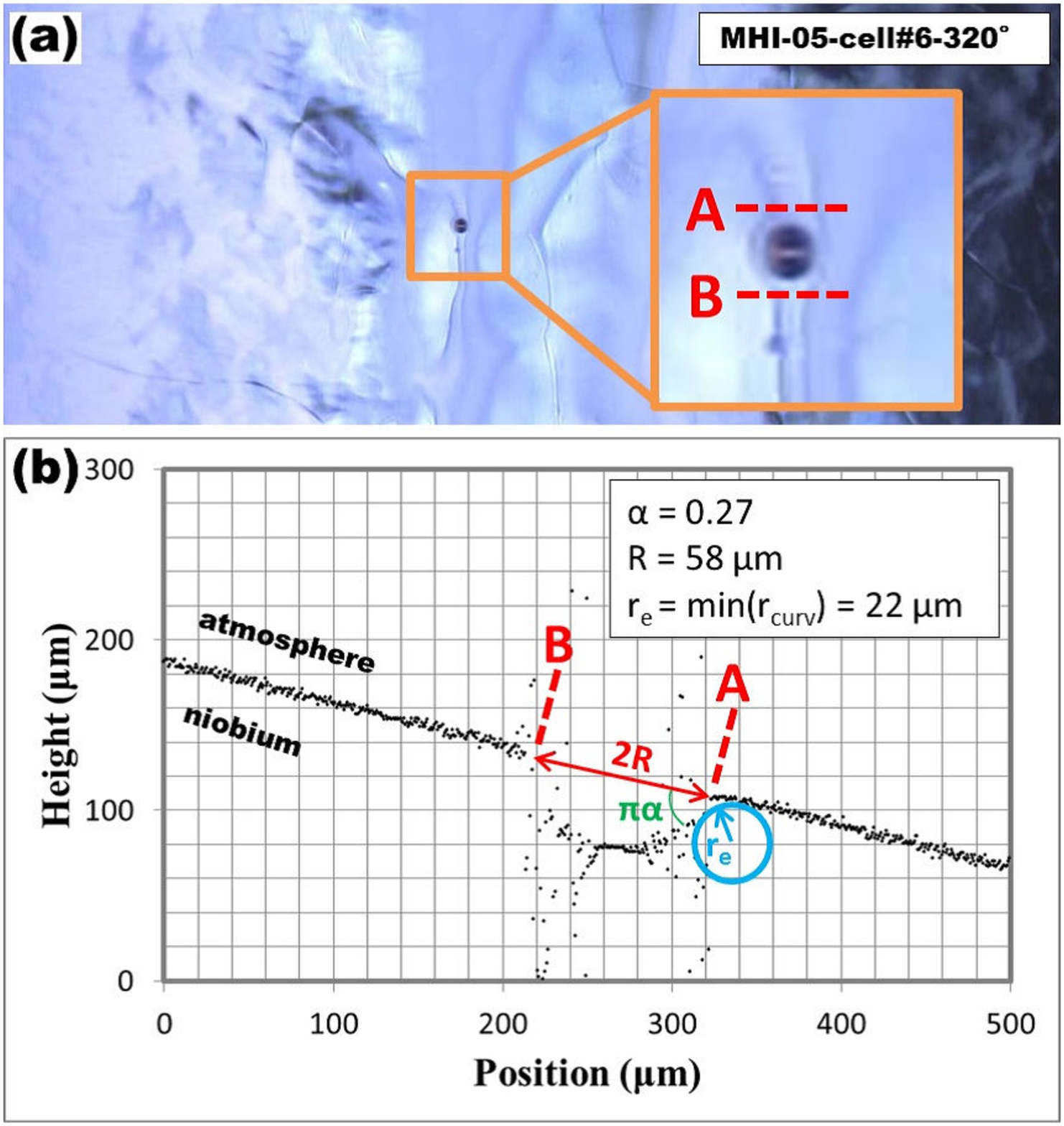}
   \end{center}\vspace{-0.6cm}
   \caption{
   A pit found at the surface of MHI-05 cavity. 
   (a) An optical image of the pit~\cite{yamamoto1}. 
   (b) Results of profilometry by a laser microscope.  
   }\label{figure8}
\end{figure}

\begin{figure}[t]
   \begin{center}
   \includegraphics[width=0.9\linewidth]{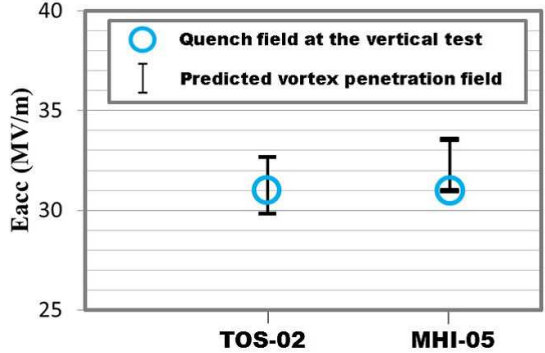}
   \end{center}\vspace{-0.6cm}
   \caption{
Comparisons between the prediction of the triangular-pit model and the vertical test results. 
The black bars correspond to the model-predictions, where $10 \% $ error of measured values of $R$, $r_e$, and $\alpha$ are assumed. 
The blue circles correspond to the quench fields, $E_{\rm acc}^{\rm (quench)}$, observed at vertical tests~\cite{yamamoto1,yamamoto2}. 
   }\label{figure9}
\end{figure}

Fig.~\ref{figure7} and Fig.~\ref{figure8} show optical images and profiles of pits found at the surfaces of TOS-02 cavity and MHI-05 cavity. 
Measured values of $R$, $r_e$, and $\alpha$ for TOS-02 and MHI-05 are summarized in the figures,   
where $r_e$ is given by the minimum value of the curvature radius $r_{\rm curv}$ calculated along the profile. 
By using these parameters, the accelerating field at which vortices start to penetrate, $E_{\rm acc}^{\rm (pen)}$, can be calculated from Eq.~(\ref{eq:Epen}). 
Fig.~\ref{figure9} shows $E_{\rm acc}^{\rm (pen)}$ calculated from Eq.~(\ref{eq:Epen}) and $E_{\rm acc}^{\rm (quench)}$ observed at vertical tests~\cite{yamamoto1,yamamoto2}. 
The agreements are remarkable, but the statistics is too small to conclude the effectiveness of the model. 
To accumulate statistics is a future work.

\section{Summary}

A model of pit should include parameters that characterize a geometry of edges of pit, where a shape of the bottom of the pit is not important. 
The simplest model of the pit is the two-dimensional pit with a triangular section. 
The well-like pit is a special case of this model. 
In this paper, MFE factor of these models were evaluated. 
The famous results for the well-like pit by Shemelin and Padamsee were reproduced analytically as an instructive excercise. 
The practically important triangular-pit model were discussed in detail, where the improved formula for the MFE factor and its predictions were shown.  
Comparisons between the prediction of the triangular-pit model and the vertical test results were also shown. 
The agreements are remarkable, but the statistics is too small to conclude the effectiveness of the model. 
To accumulate statistics is a future work.

\section*{Acknowledgements}

The author would like to express my gratitude to Yasuchika Yamamoto for his providing the experimental data. 
The author thanks Takayuki Saeki and Kensei Umemori for their encouragements and fruitful discussions.

\end{document}